\documentclass[aps,notitlepage,twocolumn]{revtex4}

\usepackage{graphicx}

\usepackage{amsmath}

\usepackage{amsfonts,amssymb}

\usepackage[utf8]{inputenc}

\usepackage{pstricks}

\usepackage{color}

\newcommand{\pslash}{\not{\hbox{\kern-2.3pt $p$}}}

\newcommand{\pdslash}{\not{\hbox{\kern-2pt $\partial$}}}

\begin{document}

\title{Production and absorption of exotic bottomonium-like states in high energy heavy ion collisions}

\author{L. M. Abreu$^{a}$, K. P. Khemchandani$^{b}$, A. Mart\'{i}nez Torres$^{c}$, F. S. Navarra$^{c}$, M. Nielsen$^{c}$ and A. L. Vasconcellos$^{a}$}

\affiliation{$^{a}$Instituto de F\'{i}sica, Universidade Federal da Bahia, Campus Universit\'{a}rio de Ondina, 40170-115, Salvador, Bahia, Brazil}

\affiliation{$^{b}$Faculdade de Tecnologia, Universidade do Estado do Rio de Janeiro, Rod. Presidente Dutra Km 298, P\'{o}lo Industrial, 27537-000, Resende, RJ, Brazil}

\affiliation{$^{c}$Instituto de  F\'{i}sica, Universidade de São Paulo, C.P. 66318, 05389-970 São Paulo, SP, Brazil}

\begin{abstract}

We investigate the production and absorption of $Z_b (10610)$ and $Z_b ^{\prime}(10650)$ states in a hadronic medium, via the processes 
 $\bar{B}^{(*)} B^{(*)} \rightarrow \pi Z_b ^{(\prime)} $ and the corresponding inverses reactions.  We use effective field Lagrangians based on an $SU(4)$-extension of the hidden gauge formalism to account for the couplings between light and heavy mesons, and a phenomenological Lagrangian involving the $ B^{\ast} B^{(\ast)} Z_b ^{(\prime)}$ vertices. The absorption cross sections are found to be much larger than the production ones. 

\end{abstract}

\maketitle




\section{Introduction}




In recent years we have witnessed the discovery of many new states,  indicating that the  heavy-hadron spectrum is much richer than expected 
in conventional constituent quark models. The benchmark in this new era of spectroscopy was the discovery of the state $X(3872)$ in 2003 by the Belle 
Collaboration \cite{Choi:2003ue}. Since then, more than twenty candidates of exotic hadron states have been observed by several collaborations. 
For a review, see references \cite{Brambilla,PDG,Chen,Hosaka}.

Among these many states, we find two charged bottomonium-like resonances,  $Z_b ^{\pm}(10610)$ and $Z_b ^{\prime \pm} (10650)$ (denoted hereafter 
as $Z_b^{\pm}$ and $Z_b ^{\prime \pm}$), observed in the processes $\Upsilon (5 S) \rightarrow \Upsilon (n S) \pi ^{\pm}\;(n=1, 2, 3)$ and 
$\Upsilon(5 S) \rightarrow \pi ^{\pm} h_b (m S)\;(m = 1, 2)$ \cite{Adachi, Bondar}. The reported masses and decay widths averaged over the mentioned 
channels are $m_{Z_b^{\pm }} = 10607.2 \pm 2.0$ MeV, $\Gamma _{Z_b^{\pm } } = 18.4 \pm 2.4$ MeV and $m_{Z_{b}^{\prime \pm }} = 10652.2 \pm 1.5$ MeV, 
$\Gamma _{Z_{b}^{\prime \pm}} = 11.5 \pm 2.2$ MeV~\cite{PDG}. Due to their charged nature and favored quantum numbers  $(I^{G}(J^P) = 1^+(1^+))$, they 
cannot be pure $b\bar b$ states and must contain at least  four quarks. Another relevant property is that, similarly to other exotic states, 
they are close to thresholds of heavy-meson bound states: $Z_b$ and $Z^{\prime}_b$ are near $B\bar B^{*}$ and $B^{*}\bar B^{*}$ thresholds, respectively. 
Thus, a natural interpretation extensively used is to suppose that they are $S$-wave deuteron-like molecules of bottomed mesons~
\cite{Bondar2,ClevenEPJ,Voloshin,Nieves2,Zhang,Sun,Yang,Ohkoda1,Ohkoda2,Li1,Li2,Li3,ClevenPRD,Ohkoda3,Dias,Kang,Huo}. Accordingly, we assume here that 
the components of $Z_b$ and $Z^{\prime}_b$ are $S$-wave molecular states of 
$\frac{1}{\sqrt{2}}(B\bar B^{*} - B^{*}\bar B) ({}^3S_1)$ and $B^*\bar B^{*}({}^3S_1)$, respectively~\cite{Ohkoda1,Ohkoda3}.

Although plausible, the meson molecule interpretation of these exotic bottomonium states is not yet firmly established. It can be argued that, due to the larger masses, 
these multiquark states should be more compact and a tetraquark configuration, i.e. two quarks and two antiquarks in a compact ``bag'',  should be favored.

In order to arrive at a consistent picture of these states, we must take advantage of all the experimental information already existent and still to be obtained. We have already 
data on the $Z_b$ and $Z^{'}_b$ masses and decay widths coming from $e^+ e^-$ collisions. More information can be obtained from the hadron colliders, in particular from the 
production cross  section  measured in proton-proton, proton-nucleus and nucleus-nucleus collisions. In the case of the much more investigated $X(3872)$, the attempt to explain  the measured  production cross section in proton-proton collisions led to the conclusion that it is very difficult to understand this state as a meson   
molecule. According to the calculations presented in \cite{han} the $X$ can be better understood as a mixture with both a molecular and a $c \bar{c}$ component. It will be interesting 
to see if the same conclusion holds for the  $Z_b$ and $Z^{'}_b$.

The experimental study of $X(3872)$ production in hadron colliders (already started \cite{hadrocoll}) and  in heavy ion collisions (HICs) will complement the 
accumulated information and help in 
discriminating between different pictures of the state. 
The same can be said about the $Z_b$ and $Z^{'}_b$  states discussed above. The advantage of working with heavy ions is that we have a much higher production rate of heavy quarks. Moreover 
in HICs there is  a quark gluon plasma (QGP) phase, where the quarks can move freely and form more easily multiquark states, specially in the hadronization transition.  
The disadvantage is that it is more difficult to identify these states experimentally, in the middle of an extremely large number of produced particles. Another disadvantage 
is that in HICs there are a number of effects and possibilities which have to be considered, for which the theoretical treatment  is still incomplete. In this work we concentrate 
on one of such aspects: the interaction of these multiquark states (more specifically of the  $Z_b$ and $Z^{'}_b$ ) with the light particles forming the hot hadronic medium which  is produced after 
the cooling and hadronization of the QGP. We will follow closely and extend the previous  works on the subject, where the interactions of the $X(3872)$ were addressed~\cite{XProd}.

After being produced at the end of the quark gluon plasma phase, the   $Z_b$ and $Z^{'}_b$  interact  with other hadrons during the expansion of the hadronic matter. Therefore, 
they can be destroyed in collisions with the comoving light mesons, but they can also be produced through the inverse reactions~
\cite{Braaten3,Artoisenet,Esposito2013,ChoLee,Guerrieri,XProd,XHMET,XHMET2}.  Since the  cross sections depend on the spatial configuration of  
these states, the strength of  these interactions depends ultimately on the internal structure of the  $Z_b$ and $Z^{'}_b$    and the measurement  of their  multiplicity 
would be very useful to  determine their structure.

Inspired by evaluations of the $X(3872)$ abundance mentioned above, in this work we study the  interactions between $Z_b$ and $Z^{\prime}_b$ and light hadrons. More precisely,   
we consider the production of $Z_b$ and $Z^{\prime}_b$ through the processes $ \bar{B} B \rightarrow \pi Z_b ^{(\prime) } $, $\bar{B}^* B \rightarrow \pi Z^{(\prime)}_b $ and 
$\bar{B}^* B^* \rightarrow \pi Z^{(\prime)}_b$ and absorption of these exotic states through the inverse processes $ \pi Z_b ^{(\prime)} \rightarrow \bar{B} B$, 
$\pi Z^{(\prime)}_b \rightarrow \bar{B}^* B $ and $ \pi Z^{(\prime)}_b\rightarrow \bar{B}^* B^* $. We obtain the amplitudes and cross sections related to these processes for $ Z^{(\prime)+}_b$ within 
the framework of $SU(4)$ effective Lagrangians~\cite{Dias,XProd}. Also, following Refs. \cite{ClevenEPJ,Ohkoda3,Huo}, we assume that the $Z_b ^+$ 
couples to the components $(\bar{B}^0 B^{\ast +} + B^{+} \bar{B}^{\ast 0} )$, while the $Z_b ^{\prime +} $ only couples to  $(B^{\ast +} \bar{B}^{\ast 0} )$.

The paper is organized as follows. In Section~\ref{Formalism} we describe the formalism, and determine the production and absorption 
amplitudes and cross sections. Then, in Section~\ref{Results} we present and discuss our results. Finally, in Section~\ref{Conclusions} we draw the concluding remarks.




\section{Formalism }

\label{Formalism}




The analysis of the processes involving the $Z^{(\prime)}_b$ production and absorption will be done in the  effective field theory approach. Accordingly, 
the Lagrangians carrying the couplings between light- and heavy-meson fields are built within the framework of an $SU(4)$-extension of the hidden gauge formalism: it consists 
of an effective theory in which the vector mesons are identified as the dynamical gauge bosons of the hidden $U(3)_V$ local symmetry in the $U(3)_L\times U(3)_R/U(3)_V$ 
non-linear sigma model \cite{XProd, Bando, Bando1, Meissner, Harada,OsetD80.114013}. The Lagrangians are given by
\begin{eqnarray}
\mathcal{L}_{PPV} & = & -ig_{PPV}\langle V^\mu[P,\partial_\mu P]\rangle ,  \nonumber \\
\mathcal{L}_{VVP} & = & \frac{g_{VVP}}{\sqrt{2}}\epsilon^{\mu\nu\alpha\beta}\langle \partial_\mu V_\nu \partial_\alpha
V_\beta P \rangle ,  
\label{eq:1}
\end{eqnarray}
where $PPV$ and $VVP$ denote pseudoscalar-pseudoscalar-vector and vector-vector-pseudoscalar vertices, respectively;
the symbol $\langle \ldots \rangle$ stands for the trace over $SU(4)$-matrices; $V_\mu$ represents a $SU(4)$ matrix, which is parametrized by 16 vector-meson fields including the 15-plet and singlet of $SU(4)$, 
\begin{eqnarray}
V_\mu = \begin{pmatrix}
	\frac{\omega}{\sqrt{2}}+\frac{\rho^0}{\sqrt{2}} & \rho^+ & K^{*+} & \bar B^{*0} \\
	\rho^{-} & \frac{\omega}{\sqrt{2}}-\frac{\rho^0}{\sqrt{2}} & K^{*0} & B^{*-} \\
	K^{*-} & \bar K^{*0} & \phi & B^{*-}_s \\
	B^{*0} & B^{*+} & B^{*+}_s & \Upsilon 
\end{pmatrix}_\mu ;
\label{eq:2}
\end{eqnarray}
$P$ is a matrix containing the 15-plet of the pseudoscalar meson fields, written in the physical basis in which $\eta$, $\eta ^{\prime}$ mixing is taken into account,
\begin{eqnarray}
 P = \begin{pmatrix}
	\frac{\eta}{\sqrt{3}}+\frac{\eta^{\prime}}{\sqrt{6}}+\frac{\pi^0}{\sqrt{2}} & \pi^{+} & K^{+} & \bar B^{0} \\
	\pi^{-} & \frac{\eta}{\sqrt{3}}+\frac{\eta^{\prime}}{\sqrt{6}}-\frac{\pi^0}{\sqrt{2}} & K^{0} & B^{-} \\
	K^{*-} & \bar K^{*0} & -\frac{\eta}{\sqrt{3}}+\sqrt{\frac{2}{3}}\eta^{\prime} & B^{-}_s \\
	B^{0} & B^{+} & B^{+}_s & \eta_b 
\end{pmatrix} . \nonumber 
\end{eqnarray}
The coupling constants $g_{PPV}$ and $g_{VVP}$ in Eq.~(\ref{eq:1}) are related to pseudoscalar-pseudoscalar-vector and vector-vector-pseudoscalar vertices, respectively, 
and are given by~\cite{XProd},
\begin{eqnarray}
g_{PPV} = \frac{m_V}{2f_\pi},\,\,\,\,\, g_{VVP} = \frac{3m^2_V}{16\pi^2f^3_\pi}
\label{eq:CouplingConstants}
\end{eqnarray} 
with $m_V$ being the mass of  the  vector meson, which we take as the mass of the $\rho$ meson, and $f_\pi$ is the pion decay constant. The coupling $g_{PPV}$ is the strong 
coupling of the $B^*$ meson to $B\pi$. Noticing that the decay $B^* \to B\pi$ is kinematically forbidden, it is not possible to determine $g_{PPV}$ from experiments. 
We then use the experimental information from the charm sector and from heavy quark symmetry \cite{XProd}, which engenders an effective $g_{PPV}$ for the vertices 
involving $B$ and $B^{*}$ mesons as 
\begin{eqnarray}
g_{PPV} = \frac{m_V}{2f_\pi}\frac{m_{B^*}}{m_{K^*}}.
\label{eq:gPPVcorrection}
\end{eqnarray}
The $m_{B^*}/m_{K^*}$ factor present in the above coupling has its origin in the heavy quark symmetry (as in Ref.~\cite{XProd}), with which the $D^* \rightarrow D \pi$ width is correctly reproduced. It must be added that our $PPV$ coupling also coincides with the value used in Ref.~\cite{Ohkoda3} where the same is determined using the heavy quark symmetry. Further, the same $PPV$ coupling has been used in Ref.~\cite{Eoepjc} where $\rho-B$ and $\rho-B^*$ interactions are studied. A comparison of the $PPV$ coupling in Ref.~\cite{Eoepjc} with the value obtained in Ref.~\cite{Flynn} within a lattice simulation shows that the two values are compatible. It is also worth to mention that our $VVP$ coupling is also comparable with the value obtained within the heavy quark symmetry in Ref.~\cite{Ohkoda3}. This is so because we only make use of one aspect of $SU(4)$, which is the connection between the different coupling constants. As shown in Ref.~\cite{Bracco}, without using any equality of masses for the light and $c$ quarks, for the case of the charm sector (where one could think that $SU(4)$ would give meaningless coupling constants) wherever the connection between couplings can be tested or compared with other approaches, the corresponding $SU(4)$ relations give very similar results.

Next, we can introduce the couplings of the $Z^{(\prime)}_b$ to $B^{(\ast)}$-meson fields. We emphasize that in the present approach we treat  $Z^{(\prime)}_b$ as an elementary 
degree of freedom, with quantum numbers $J^{P}= 1^{+}$. Also, following Refs.~\cite{ClevenEPJ,ClevenPRD,Huo}, we assume that 
$Z_b ^+(10610)$ couples to the components $(\bar{B}^0 B^{\ast +} + B^{+} \bar{B}^{\ast 0} )$, while $Z_b ^{\prime +} (10650)$ couples only to the channel $(B^{\ast +} 
\bar{B}^{\ast 0} )$. Then, the phenomenological Lagrangians involving the $Z^{(\prime)}_b B^{\ast} B^{(\ast)}$ vertices are
\begin{eqnarray}
\mathcal{L}_{ZBB^*}  & = & g_{ZBB^*} (B Z ^{ \mu} B^{*  \dagger}_\mu + B^{*}_\mu Z ^{ \mu} B^{\dagger}), \nonumber \\
\mathcal{L}_{Z^{\prime}B^*B^*} & = & ig_{Z^{\prime}B^*B^*}\epsilon^{\alpha\beta \mu\nu} B^{*}_\alpha \partial_{\beta} Z^{\prime}_{\mu}  B^{*\dagger}_{\nu},
\label{eq:3}
\end{eqnarray}
where $g_{ZBB^*} $ and $g_{Z^{\prime}B^*B^*} $ are the coupling constants of the $ZBB^*$ and $Z^{\prime}B^*B^*$ vertices, respectively; $B = B^{-}$ , $B_{\mu} ^{\ast} = B_{\mu} ^{\ast -}$ and  $Z_{\mu} ^{(\prime)} = Z_{\mu} ^{(\prime) -}$ and the greek letters indicate Lorentz indices.

\begin{figure}[!ht]
    \centering
        \includegraphics[width=8.0cm]{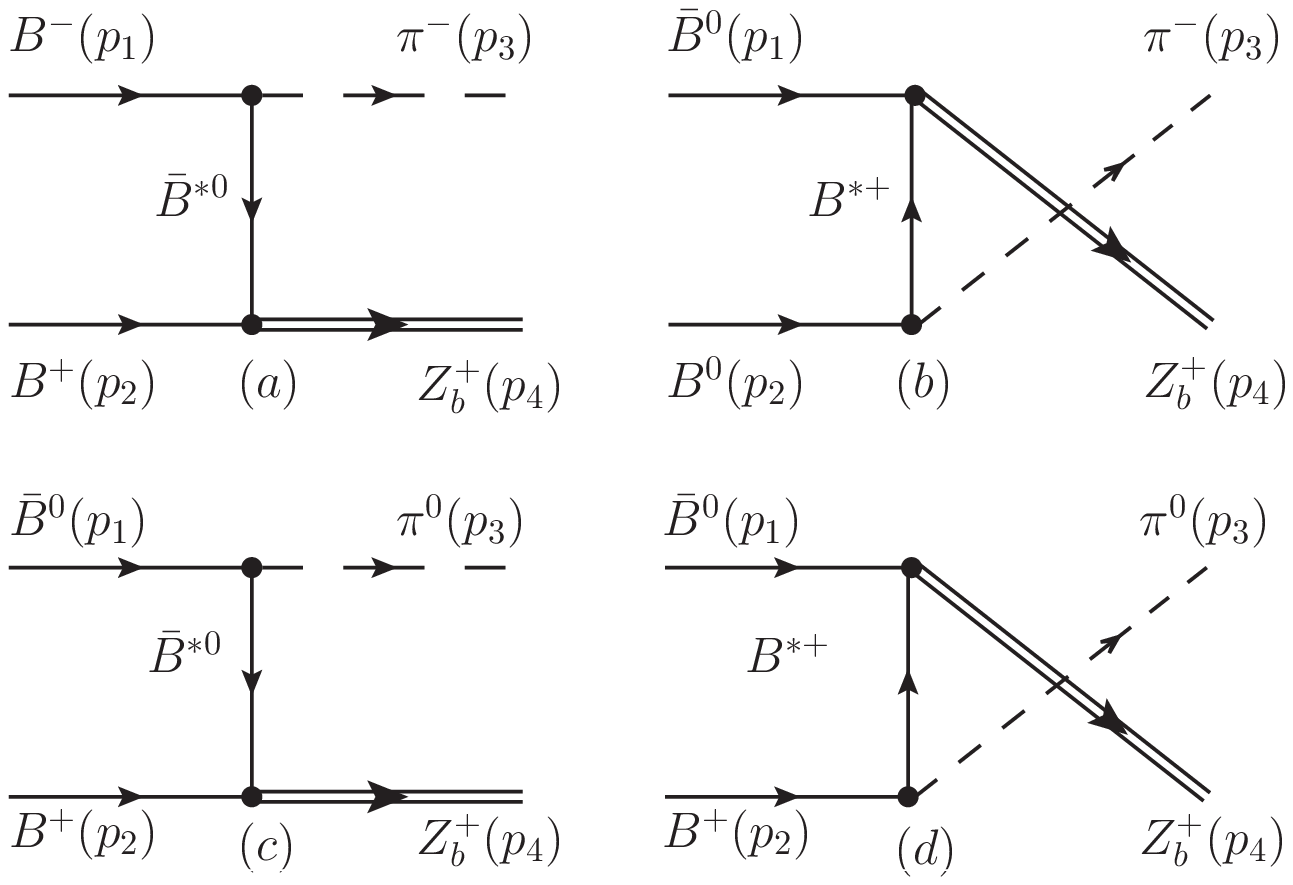}
        \caption{Diagrams contributing to the process $ \bar{B} B \rightarrow \pi Z_b $.}
\label{FIG1}
\end{figure}

\begin{figure}[!ht]
    \centering
        \includegraphics[width=8.0cm]{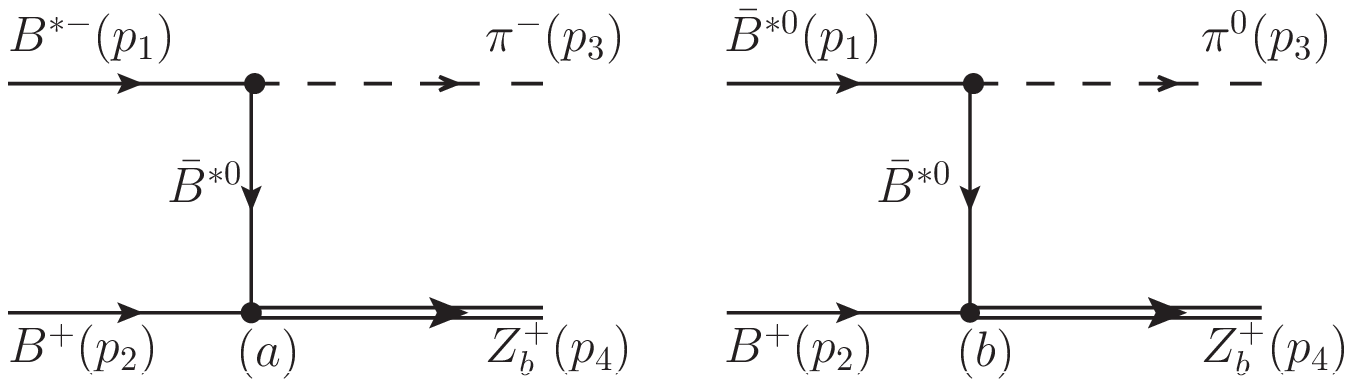}
        \caption{Diagrams contributing to the process $ \bar{B}^* B \rightarrow \pi Z_b $.}
    \label{FIG2}
\end{figure}

\begin{figure}[!ht]
    \centering
        \includegraphics[width=8.0cm]{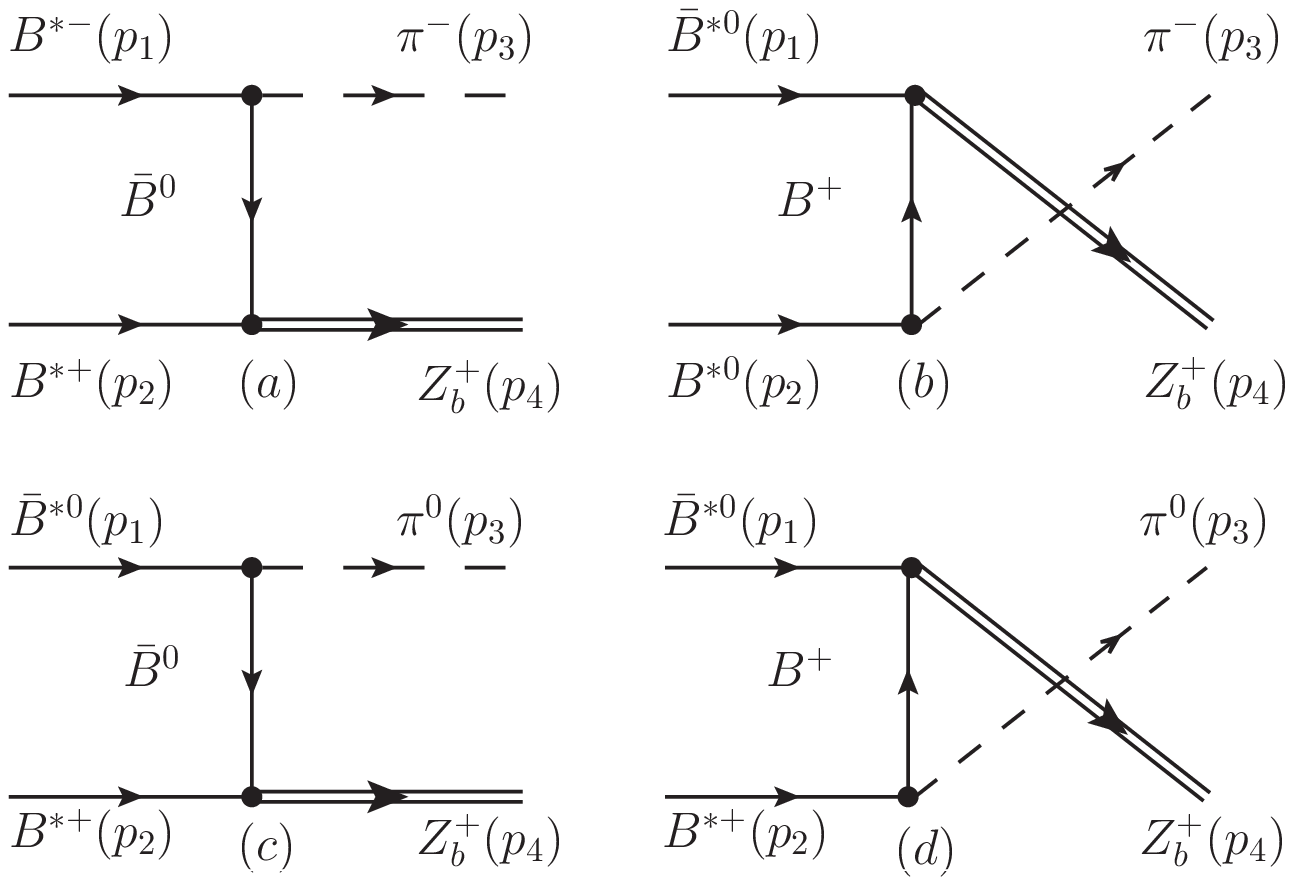}
        \caption{Diagrams contributing to the process $ \bar{B}^* B^* \rightarrow \pi Z_b $.}
    \label{FIG3}
\end{figure}

\begin{figure}[!ht]
    \centering
        \includegraphics[width=8.0cm]{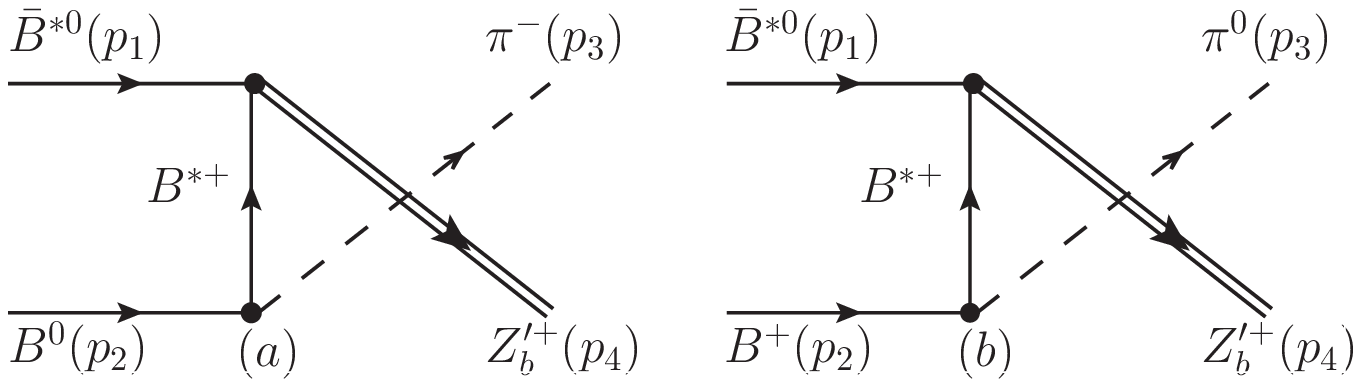}
        \caption{Diagrams contributing to the process $ \bar{B}^* B \rightarrow \pi Z_b ^{\prime} $.}
    \label{FIG2Prime}
\end{figure}

\begin{figure}[!ht]
    \centering
        \includegraphics[width=8.0cm]{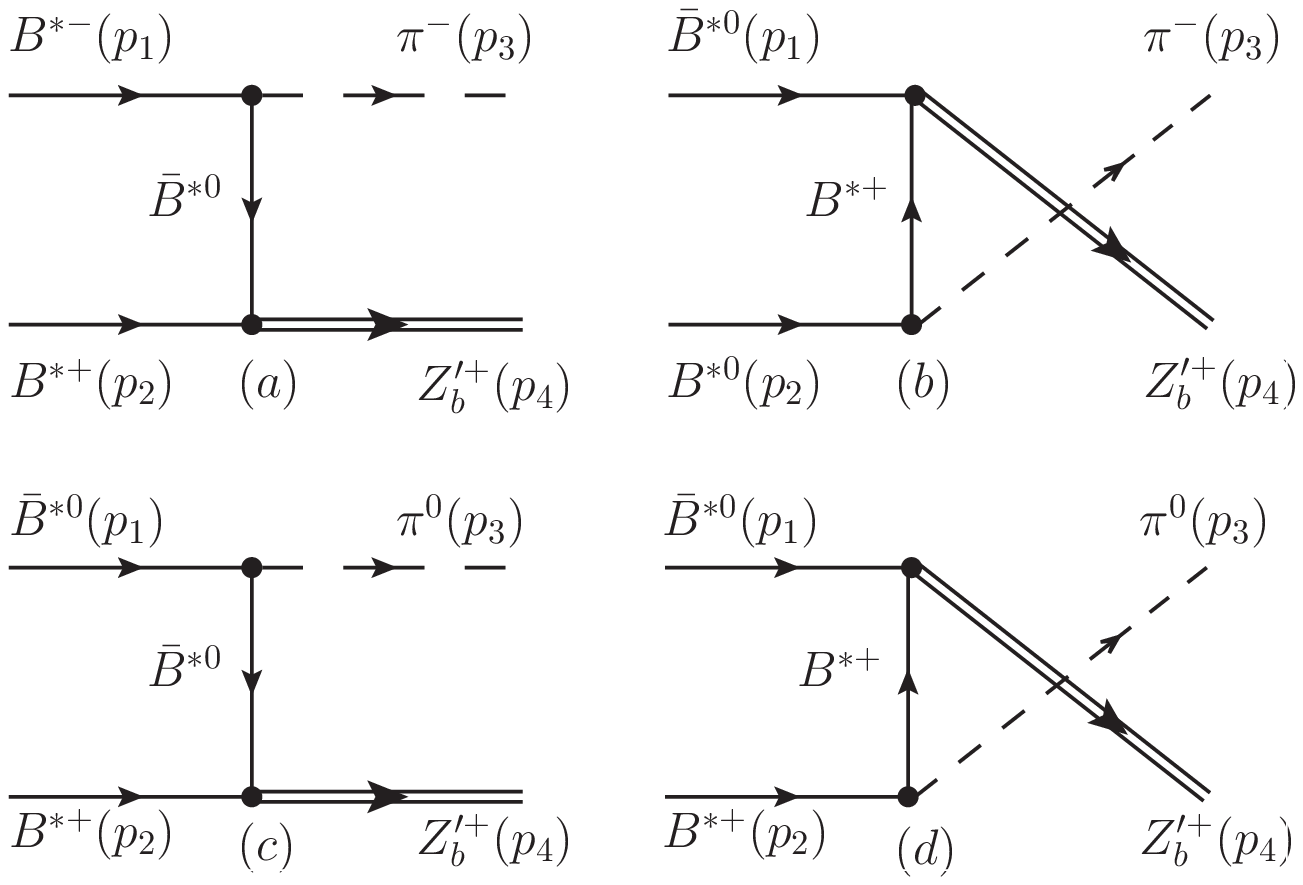}
        \caption{Diagrams contributing to the process $ \bar{B}^* B^* \rightarrow \pi Z_b ^{\prime} $.}
    \label{FIG3Prime}
\end{figure}

Now we can determine the transition amplitudes for the processes $\bar{B} B \rightarrow \pi Z_b ^{(\prime) } $, $\bar{B}^* B \rightarrow \pi Z^{(\prime)}_b $ and 
$\bar{B}^* B^* \rightarrow \pi Z^{(\prime)}_b$, by using the Lagrangians in Eqs.~(\ref{eq:1}) and (\ref{eq:3}). In Figs.~\ref{FIG1}-\ref{FIG3Prime} we show the diagrams 
associated to the  mentioned processes at leading order, with the specification of the charges of the incoming bottomed mesons and of the particles in the  final state, keeping in mind 
that the diagrams in Figs. \ref{FIG1}, \ref{FIG2} and \ref{FIG3} are related to the processes involving $Z^{+}_b$ production, while diagrams in Figs. \ref{FIG2Prime} and \ref{FIG3Prime} to the $Z^{\prime +}_b$.

The amplitudes associated to the $t$-channel diagrams shown in Figs.~\ref{FIG1}-\ref{FIG3} for the  $ \bar{B} B , \bar{B}^* B , \bar{B}^* B^* \rightarrow \pi Z_b $ 
processes are, respectively,  
\begin{eqnarray}
\mathcal{T}^{(Q_{1i},Q_{2i})}_1 & = & T^{(Q_{1i},Q_{2i})}_1 g_{PPV}g_{ZBB^*}\frac{1}{t-m^2_{\bar{B}^*}}\nonumber \\
 && \times \left[(p_1 + p_3)_\mu + \frac{m^2_B-m^2_\pi}{m^2_{B^{*}}}p_{2\mu}\right]\epsilon^{*\mu}_Z(p_4); \nonumber \\
\mathcal{T}^{(Q_{1i},Q_{2i})}_2 & = & T^{(Q_{1i},Q_{2i})}_2 g_{VVP}g_{ZBB^*} \frac{1}{t-m^2_{\bar{B}^*}} \nonumber \\
&& \times \epsilon^{\mu\nu\alpha\beta} p_{1\mu}p_{3\alpha}\epsilon_{B^{*}\nu}(p_1)\epsilon^{*}_{Z\beta}(p_4);\nonumber \\
\mathcal{T}^{(Q_{1i},Q_{2i})}_3 & = & T^{(Q_{1i},Q_{2i})}_3 g_{PPV}g_{ZBB^*} \frac{1}{t-m^2_{\bar{B}}} \nonumber \\
&& \times  p_{3\mu}\epsilon^{\mu}_{B^*}(p_1) \epsilon^{\sigma}_{B^{*}}(p_2)\epsilon^{*}_{Z\sigma}(p_4);
\label{eq:tchannelZproduction}
\end{eqnarray}
while the amplitudes associated to the $u$-channel diagrams shown in Figs.\ref{FIG1} and \ref{FIG3} are, respectively, 
\begin{eqnarray}
\mathcal{U}^{(Q_{1i},Q_{2i})}_1 & = & U^{(Q_{1i},Q_{2i})}_1 g_{PPV}g_{ZBB^*}\frac{1}{u-m^2_{\bar{B}^*}}\nonumber \\
 & & \times \left[(p_2 + p_3)_\mu + \frac{m^2_B - m^2_\pi}{m^2_{B^{*}}}p_{1\mu}\right]\epsilon^{*\mu}_Z(p_4); \nonumber \\
%
%
%
\mathcal{U}^{(Q_{1i},Q_{2i})}_3 & = & U^{(Q_{1i},Q_{2i})}_3 g_{PPV}g_{ZBB^*} \frac{1}{u-m^2_{\bar{B}}} \nonumber \\
&& \times p_{3\mu}\epsilon^{\mu}_{B^*}(p_2) \epsilon^{\sigma}_{B^{*}}(p_1)\epsilon^{*}_{Z\sigma}(p_4).
\label{eq:uchannelZproduction}
\end{eqnarray}
The quantities $ T^{(Q_{1i},Q_{2i})}_{r}$ and $U^{(Q_{1i},Q_{2i})}_{r} \,\,(r=1,\cdots,3)$  appearing in Eqs. (\ref{eq:tchannelZproduction}) and 
(\ref{eq:uchannelZproduction}) are isospin coefficients of the scattering amplitudes for $t$ and $u$-channels, respectively, and are defined in 
Table \ref{table1}; $(Q_{1i},Q_{2i})$ denotes the charges of particles in the initial state; $p_1$, $p_2$ ($p_3$, $p_4$) represent the momentum of the particles in the initial 
(final) states; $m_{B}$, $m_{B^{\ast}}$, $m_{\bar{B}}$, $m_{\bar{B}^{\ast}}$ and $m_{\pi}$ are average masses of the $B$, $B^{\ast}$, $\bar{B}$, $\bar{B}^{\ast}$ and 
$\pi$ mesons; $\epsilon_{B^{*}}(p_1) $ and $\epsilon^{*}_{{Z}^{(\prime) }}(p_4)$ are the polarization vectors of $B^{\ast}$ mesons and $Z^{(\prime) }_b $ states, 
respectively.



In the case of $Z_b^{\prime +}$ production, we note that there is no diagram contributing to  $ \bar{B} B  \rightarrow \pi Z_b ^{\prime} $ reaction,  since there is no 
$ B \bar{B}^{\ast}  Z_b ^{\prime} $ vertex. Thus, the amplitudes related to the $t$-channel diagrams shown in Fig.~\ref{FIG3Prime} for the 
$ \bar{B}^* B^* \rightarrow \pi Z_b^{\prime} $ process are,
\begin{eqnarray}
\mathcal{T}^{(Q_{1i},Q_{2i})}_5 & = & T^{(Q_{1i},Q_{2i})}_5 g_{VVP}g_{Z'B^*B^*}\frac{1}{t-m^2_{\bar{B}^*}} \nonumber \\
&& \times \epsilon^{\mu\nu\alpha\beta}\epsilon^{\lambda\sigma\gamma}_{\hspace{0.45cm}\beta} p_{1\lambda}p_{3\gamma}p_{4\mu} \nonumber \\
&& \times \epsilon_{B^*\sigma}(p_1)\epsilon_{B^* \alpha}(p_2)\epsilon^{*}_{Z'\nu}(p_4);
\label{eq:tchannelZPrimeproduction}
\end{eqnarray}
while the amplitudes for the $u$-channel diagrams in Figs.~\ref{FIG2Prime} and \ref{FIG3Prime} associated to the $ \bar{B}^* B , \bar{B}^* B^* \rightarrow \pi Z_b ^{\prime} $ 
processes are, respectively,
\begin{eqnarray}
\mathcal{U}^{(Q_{1i},Q_{2i})}_4 & = & U^{(Q_{1i},Q_{2i})}_4 g_{PPV}g_{Z'B^*B^*} \frac{1}{u-m^2_{B^*}} \nonumber \\
&& \times  \epsilon^{\mu\nu\alpha\beta} p_{4\mu}\left[(p_2+p_3)_\alpha + \frac{m^2_B - m^2_\pi}{m^2_{B^*}}p_{1\alpha}\right] \nonumber \\
&& \times \epsilon_{B^{*}\beta}(p_1)\epsilon^{*}_{Z'\nu}(p_4);\nonumber \\
\mathcal{U}^{(Q_{1i},Q_{2i})}_5 & = & U^{(Q_{1i},Q_{2i})}_5 g_{VVP}g_{Z'B^*B^*}\frac{1}{u-m^2_{\bar{B}^*}} \nonumber \\
&& \times \epsilon^{\mu\nu\beta\alpha}\epsilon^{\lambda\gamma\delta}_{\hspace{0.45cm}\alpha} p_{2\gamma}p_{3\lambda}p_{4\mu} \nonumber \\ 
&& \times \epsilon_{\bar{B}^*\beta}(p_1)\epsilon_{B^* \delta}(p_2)\epsilon^{*}_{Z'\nu}(p_4). 
\label{eq:uchannelZPrimeproduction}
\end{eqnarray}
Again, the isospin coefficients $ T^{(Q_{1i},Q_{2i})}_{4}$, $U^{(Q_{1i},Q_{2i})}_{4}$ and $U^{(Q_{1i},Q_{2i})}_{4}$ are defined in Table \ref{table1}.



The scattering amplitudes associated with the inverse processes $\pi Z_b ^{(\prime) +} \rightarrow \bar{B} B, \bar{B}^* B ,  \bar{B}^* B^* $ and 
$\pi Z^{\prime +}_b \to \bar{B}^* B ,  \bar{B}^* B^* $ can be determined as above, by using the  correspondence $p_1 \leftrightarrow p_3$ and  $p_2 \leftrightarrow p_4$.


\begin{table}
\caption{Isospin coefficients $ T^{(Q_{1i},Q_{2i})}_{r}$ and $U^{(Q_{1f},Q_{2f})}_{r} \,\,(r=1,\cdots,5)$ appearing in Eqs. (\ref{eq:tchannelZproduction}) and 
(\ref{eq:uchannelZproduction}).} 
\begin{center}
\begin{tabular}{c|c|c}
\hline
\hline
Diagram & Process  & $  T^{(Q_{1i},Q_{2i})}_{r}$ or $U^{(Q_{1i},Q_{2i})}_{r}$ \\
\hline
(1a) & $B^{-} B^{+} \rightarrow \pi ^{-} Z_b ^{+}  $   & 1    \\
(1b) & $\bar{B}^{0} B^{0} \rightarrow \pi ^{-} Z_b ^{+} $  & $-1$  \\
(1c) & $\bar{B}^{0} B^{+}\rightarrow \pi ^{0} Z_b ^{+}  $  & $\frac{1}{\sqrt{2}}$ \\
(1d) & $\bar{B}^{0} B^{+} \rightarrow \pi ^{0} Z_b ^{+} $ & $ \frac{1}{\sqrt{2}}$  \\
\hline
(2a) & $B^{\ast -} B^{+} \rightarrow \pi ^{-} Z_b ^{+}  $  & $\frac{1}{\sqrt{2}}$  \\
(2b) & $\bar{B}^{\ast 0} B^{+} \rightarrow \pi ^{0} Z_b ^{+} $ & $\frac{1}{2}$ \\
\hline
%
%
%
%
(3a) & $B^{\ast -} B^{\ast +} \rightarrow \pi ^{-} Z_b ^{+}  $   & 2  \\
(3b) & $\bar{B}^{\ast 0} B^{\ast 0} \rightarrow \pi ^{-} Z_b ^{+} $ & $-2$  \\
(3c) & $\bar{B}^{\ast 0} B^{\ast +} \rightarrow \pi ^{0} Z_b ^{+}  $ & $ \sqrt{2}$ \\
(3d) & $\bar{B}^{\ast 0} B^{\ast +} \rightarrow \pi ^{0} Z_b ^{+}  $ &$ \sqrt{2}$ \\
\hline
(4a) & $\bar{B}^{\ast 0} B^{0} \rightarrow \pi ^{-} Z_b ^{\prime \, +} $ & $1$  \\
(4b) & $\bar{B}^{\ast 0} B^{+} \rightarrow \pi ^{0} Z_b ^{\prime \, +} $ & $-\frac{1}{\sqrt{2}}$ \\
\hline
(5a) & $B^{\ast -} B^{\ast +} \rightarrow \pi ^{-} Z_b ^{\prime \, +}  $ & $- \frac{1}{\sqrt{2}}$  \\
(5b) & $\bar{B}^{\ast 0} B^{\ast 0} \rightarrow \pi ^{-} Z_b ^{\prime \, +} $ & $\frac{1}{\sqrt{2}}$ \\
(5c) & $\bar{B}^{\ast 0} B^{\ast +} \rightarrow \pi ^{0} Z_b ^{\prime \, +} $ & $ -\frac{1}{2}$ \\
(5d) & $\bar{B}^{\ast 0} B^{\ast +} \rightarrow \pi ^{0} Z_b ^{\prime \, +} $ & $ -\frac{1}{2}$ \\
\hline
\hline
\end{tabular}
\end{center}
\label{table1}
\end{table}



At this point we are able to determine the isospin-spin-averaged cross section for the processes $\bar{B}B, \bar{B}^*B, \bar{B}^*B^* \to \pi Z^{(\prime) +}_b$, 
which in the center of mass (CM) frame  is defined as
\begin{eqnarray}
\sigma_r(s) = \frac{1}{64 \pi^2 s }  \frac{|\vec{p}_{f}|}{|\vec{p}_i|}  \int d \Omega \overline{\sum_{S, I}}|\mathcal{M}_r(s,\theta)|^2, 
\label{eq:CrossSection}
\end{eqnarray}
where $r = 1,2,3$ label processes associated with $Z^{+}_b$ production, and $r = 4,5$ to $Z^{\prime +}_b$ production, as in the  notation  introduced above; 
$\sqrt{s}$ is the CM energy;  $|\vec{p}_{i}|$ and $|\vec{p}_{f}|$ denote the tri-momenta of initial and final particles in the CM frame, respectively; the symbol 
$\overline{\sum_{S,I}}$ represents the sum over the spins and isospins of the particles in the initial and final state, weighted by the isospin and spin degeneracy 
factors of the two particles forming the initial state for the reaction $r$, i.e. \cite{XProd} 
\begin{eqnarray}
\overline{\sum_{S,I}}|\mathcal{M}_r|^2 & = & \frac{1}{(2I_{1i,r}+1)(2I_{2i,r}+1)} \nonumber \\
&& \times \frac{1}{(2S_{1i,r}+1)(2S_{2i,r}+1)} \sum_{S,I}|\mathcal{M}_r|^2, \nonumber \\
\label{eq:DegeneracyFactors}
\end{eqnarray}
where 
\begin{eqnarray}
\sum_{S,I} |\mathcal{M}_r|^2 = \sum_{Q_{1i}, Q_{2i}} \left[\sum_{S}\left|\mathcal{M}^{(Q_{1i},Q_{2i})}\right|^2\right].
\label{eq:Sum}
\end{eqnarray}
Notice that the charges of the two particles forming the initial state for the processes in Figs.~\ref{FIG1}-\ref{FIG3Prime} can be combined, giving a total 
charge $Q_r = Q_{1i} + Q_{2i} = 0, +1$. We have then three possibilities: $(0,0)$, $(-,+)$ and $(0,+)$, yielding
\begin{eqnarray}
\sum_{S,I} |\mathcal{M}_r|^2  = \sum_{S}\left(|\mathcal{M}^{(0,0)}_r|^2 + |\mathcal{M}^{(-,+)}_r|^2 + |\mathcal{M}^{(0,+)}_r|^2\right). \nonumber \\
\label{eq:SumSpin}
\end{eqnarray}

Each amplitude $\mathcal{M}^{(Q_{1i},Q_{2i})}_r$ in Eq. (\ref{eq:Sum}) can be written, in general, as 
\begin{eqnarray}
\mathcal{M}^{(Q_{1i},Q_{2i})}_r = \mathcal{T}^{(Q_{1i},Q_{2i})}_r +  \mathcal{U}^{(Q_{1i},Q_{2i})}_r, 
\label{eq:Scattering}
\end{eqnarray}
where $\mathcal{T}^{(Q_{1i},Q_{2i})}_r$ and $\mathcal{U}^{(Q_{1i},Q_{2i})}_r$ are the $t$- and $u$-channel amplitudes given in 
Eqs. (\ref{eq:tchannelZproduction})-(\ref{eq:uchannelZPrimeproduction}).





\section{Results}

\label{Results}





In this Section we analyze the $Z_b ^{(\prime) +}$-production cross sections as a function of CM energy $\sqrt{s}$.
The values of physical quantities and coupling constants used here are~\cite{PDG}: $m_{\pi} = 137.3$ MeV; $m_{B} = 5279.4$ MeV; 
$m_{B^{*}} = 5324.8$ MeV; $m_Z = 10607.2$ MeV; $m_{Z^{\prime}} = 10652.2$ MeV; $m_{V} \equiv m_{\rho} = 775$ MeV; and $f_{\pi} = 93$ MeV. 
As for the $g_{ZBB^*} $  and $g_{Z^{\prime}B^*B^*}$ coupling constants introduced in Eq.~(\ref{eq:3}),
the values considered here are those obtained in Ref.~\cite{Li3} (in accordance with the ones used in Ref.~\cite{Ohkoda3}): 
\begin{eqnarray}
g_{ZBB^*} & = & 13.10 _{-0.88} ^{+0.83} \,\, \mathrm{GeV}, \nonumber \\
g_{Z^{\prime}B^*B^*} & = & 1.04 _{-0.1} ^{+0.1}  \,.
\label{CouplConst}
\end{eqnarray}
To take into account the uncertainties of these couplings, the results discussed below will be represented by shaded regions in the plots.

\begin{figure}[!ht]
    \centering
        \includegraphics[{width=8.0cm}]{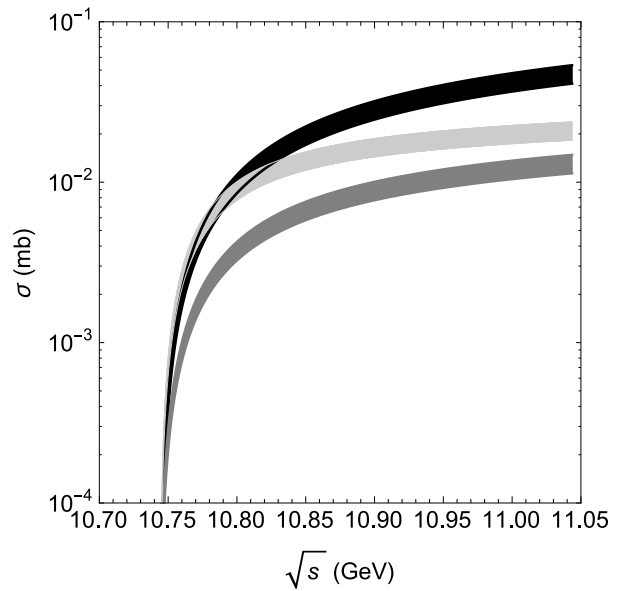} 
        \caption{
				Cross sections for the processes $\bar{B}  B  \rightarrow  \pi Z_b $ (dark shaded region), $\bar{B}^* 
B \rightarrow  \pi Z_b $ (medium shaded region) and  $\bar{B}^*  B^* \rightarrow  \pi Z_b $ (light shaded region), as function of CM energy $\sqrt{s}$.
}
    \label{FIG4}
\end{figure}

In Fig.~\ref{FIG4} the $Z_b ^+$ production cross sections  are plotted as a function of the CM energy $\sqrt{s}$. 
We see that the cross sections are  $ \sim 3 \times 10^{-3} - 5 \times  10^{-2} $mb for $ 10.80 \leq \sqrt{s} \leq 11.05 $ GeV. 
From the figure we can see that the biggest contribution to the $Z_b ^+$ production comes from the reaction  with $B \bar{B}$ in the initial state.
The $\bar{B}  \,  B  \rightarrow  \pi \,  Z_b  $ cross section is bigger than the others by a factor about 2-3.


\begin{figure}[!ht]
    \centering
        \includegraphics[{width=8.0cm}]{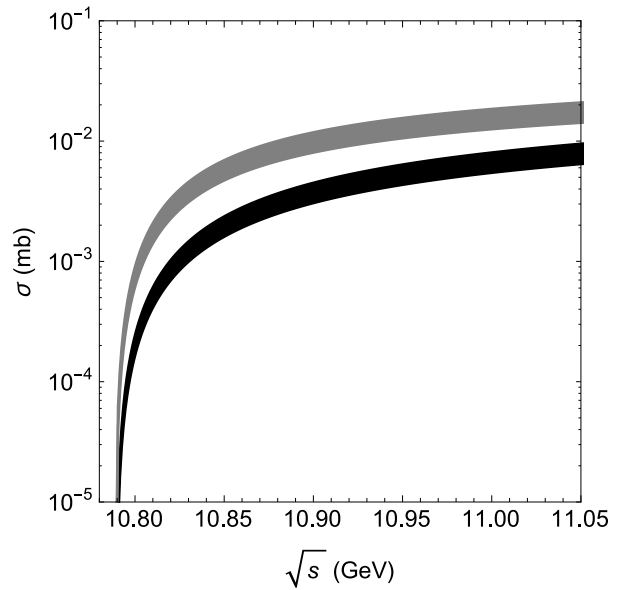} 
        \caption{
				Cross sections for the processes $\bar{B}^*  B  \rightarrow  \pi Z_b ^{\prime} $ (dark shaded region) and 
$\bar{B}^*  B^* \rightarrow  \pi Z_b ^{\prime} $ (light shaded region), as function of CM energy $\sqrt{s}$.}
    \label{FIG5}
\end{figure}

The $Z_b ^{\prime +}$ production cross sections are plotted in Fig.~\ref{FIG5} as a  function of the CM energy $\sqrt{s}$. 
Remembering that in this case there is no reaction with initial $\bar{B}  B$ state at leading order, the two relevant processes have cross sections found to be 
 $\sim 6 \times 10^{-4} - 2 \times 10^{-2} $ mb for $ 10.82 \leq \sqrt{s} \leq 11.05 $ GeV, but with the reaction with initial $\bar{B}^*  B^*$ state having the largest cross 
section by a factor about 2-3.


\begin{figure}[!ht]
    \centering
        \includegraphics[{width=8.0cm}]{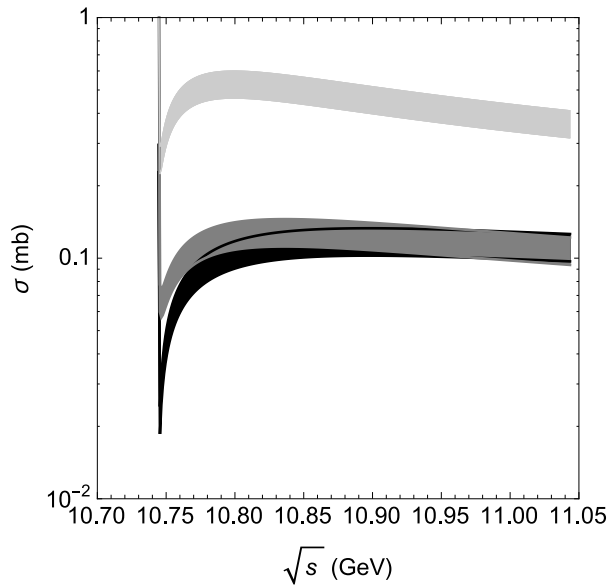} 
        \caption{
				Cross sections for the processes $  \pi Z_b  \rightarrow \bar{B}  B $ (dark shaded region), 
$\pi Z_b  \rightarrow \bar{B}^*  B  $ (medium shaded region) and  $ \pi Z_b  \rightarrow \bar{B}^*  B^* $ (light shaded region), as function of CM energy $\sqrt{s}$.}
    \label{FIG6}
\end{figure}

For completeness, the cross sections related to the inverse processes can be also analyzed. In Fig.~\ref{FIG6} the $Z_b ^+$ absorption cross sections are plotted as a function of the  CM energy $\sqrt{s}$.
They are found to be $\sim 8 \times 10^{-2} - 6 \times 10^{-1} $ mb for $ 10.80 \leq \sqrt{s} \leq 11.05 $ GeV.  As can be seen, the reaction with the  final $\bar{B}^*  B^*  $ state has cross section larger by a factor about 3-4 with respect to other reactions.

Also, another relevant point is the comparison among the $Z_b ^+$ production and absorption cross sections reported in Figs.~\ref{FIG4} and \ref{FIG6}, respectively: the $Z_b$-production cross sections are smaller than the absorption ones by a factor about 2-10, depending on the specific channel. The essence of the difference between production and absorption cross sections is due to kinematic effects.


\begin{figure}[!ht]
    \centering
        \includegraphics[{width=8.0cm}]{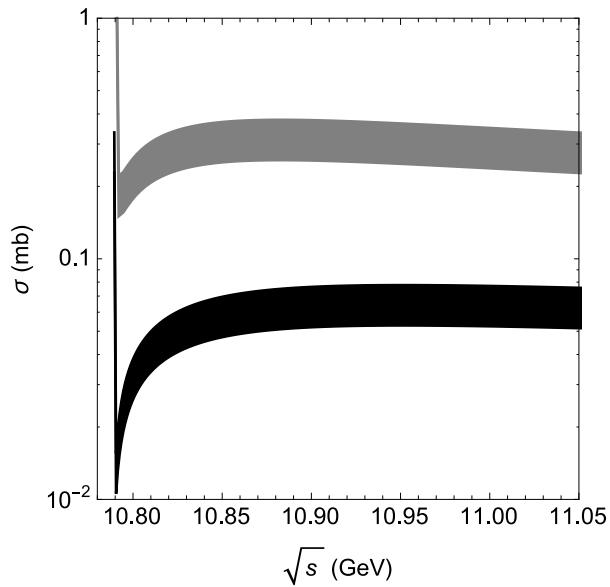} 
        \caption{
				Cross sections for the processes $\pi Z_b ^{\prime}  \rightarrow \bar{B}^*  B  $ (dark shaded region) 
and  $ \pi Z_b ^{\prime}  \rightarrow \bar{B}^*  B^* $ (light shaded region), as function of CM energy $\sqrt{s}$.}

    \label{FIG7}
\end{figure}

In Fig.~\ref{FIG7} the $Z_b ^{\prime +}$ absorption cross sections  are plotted as a function of the CM energy $\sqrt{s}$. The order of the cross sections is found to be $ 4 \times 10^{-2} - 3 \times 10^{-1} $ 
mb for $ 10.82 \leq \sqrt{s} \leq 11.05 $ GeV. The reaction with final $\bar{B}^*  B^*  $  state has the largest cross section by a factor about 2-3 with respect to reaction with final $\bar{B}^*  B  $ state.
In addition, it can be noticed that the $Z_b ^{\prime}$ absorption cross sections in Fig.~\ref{FIG5} are greater than the $Z_b ^{\prime}$  production cross sections in Fig.~\ref{FIG7} by a factor about 8-10, depending on the specific channel. This behavior is qualitatively similar to the case involving the $Z_b$ state.


The findings reported above can be compared to previous works. In particular, in Ref.~\cite{XHMET2} the $Z_b ^{(\prime) }$  production cross sections are analyzed making use of Heavy-Meson Effective Theory (HMET), taking as guiding principles chiral $SU(3)_{L} \times SU(3)_{R}$ and heavy quark symmetries. Considering the relevant scales for HMET, in this approach  $p_{\pi}$ (the tri-momentum of the pion) is requested to be much less than $\Lambda_{\chi} = 4 \pi f_{\pi} \sim 1$ GeV. This fact engenders a range of validity for the collision energy of each process. Thus, restricting the comparison to the energy ranges in which the results reported in Ref.~\cite{XHMET2} are valid, it can be noticed that $Z_b ^{(\prime) +}$  production cross sections in the present work are smaller  by a factor about 10. We believe that this discrepancy is mainly due to the difference between the magnitude of the couplings, since  in Ref.~\cite{XHMET2} the $g_{ZBB^*} $  and $g_{Z^{\prime}B^*B^*}$  coupling constants employed are larger (by a factor of $\sqrt{8}$) as compared to the ones used here.



\subsection{Inclusion of Form Factors}




\begin{figure}[!ht]
    \centering
        \includegraphics[{width=8.0cm}]{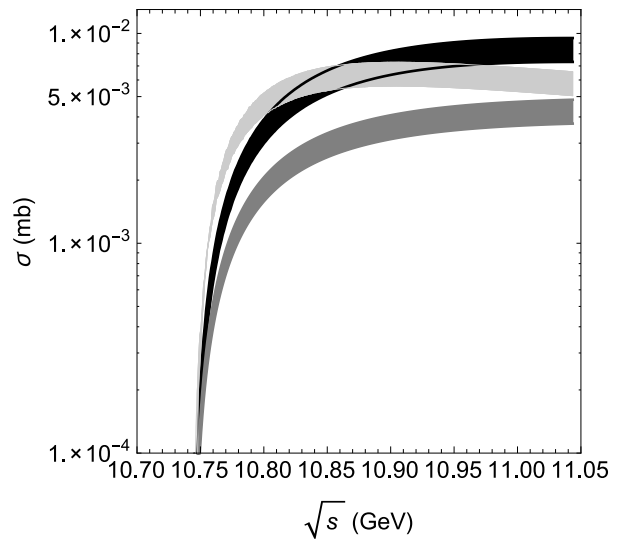} 
        \caption{Same as in Fig.~\ref{FIG4}, but with inclusion of form factors. }
    \label{FIG8}
\end{figure}



\begin{figure}[!ht]
    \centering
        \includegraphics[{width=8.0cm}]{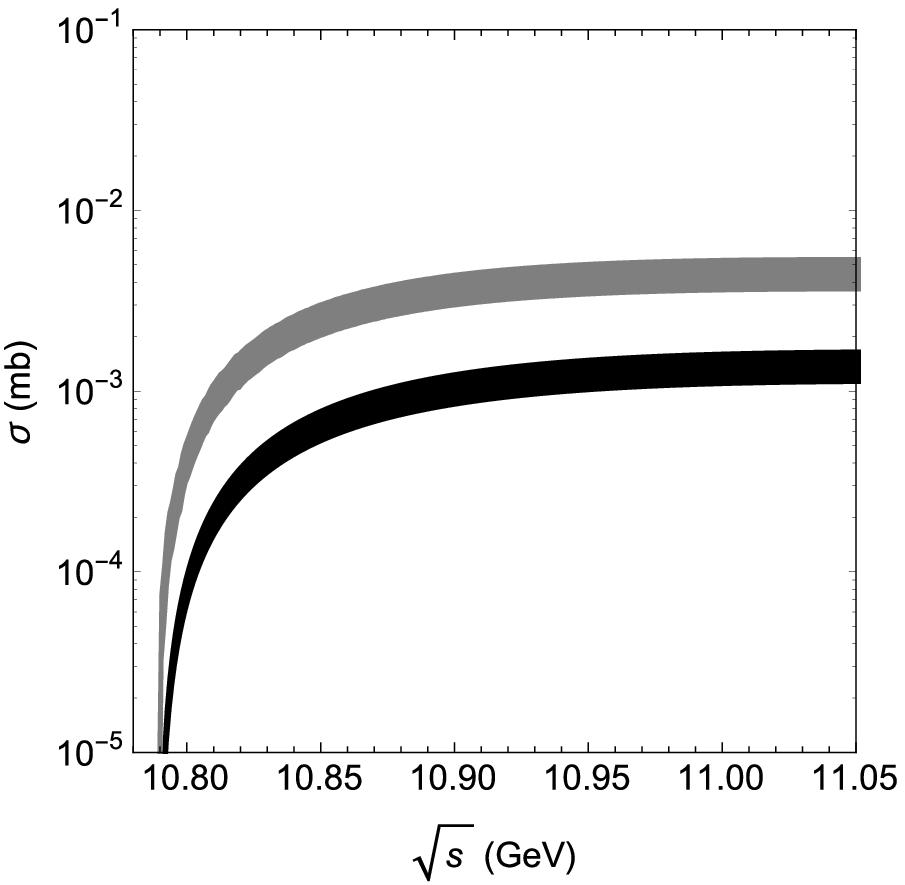} 
        \caption{Same as in Fig.~\ref{FIG5}, but with inclusion of form factors. }
    \label{FIG9}
\end{figure}



\begin{figure}[!ht]
  \centering
        \includegraphics[{width=8.0cm}]{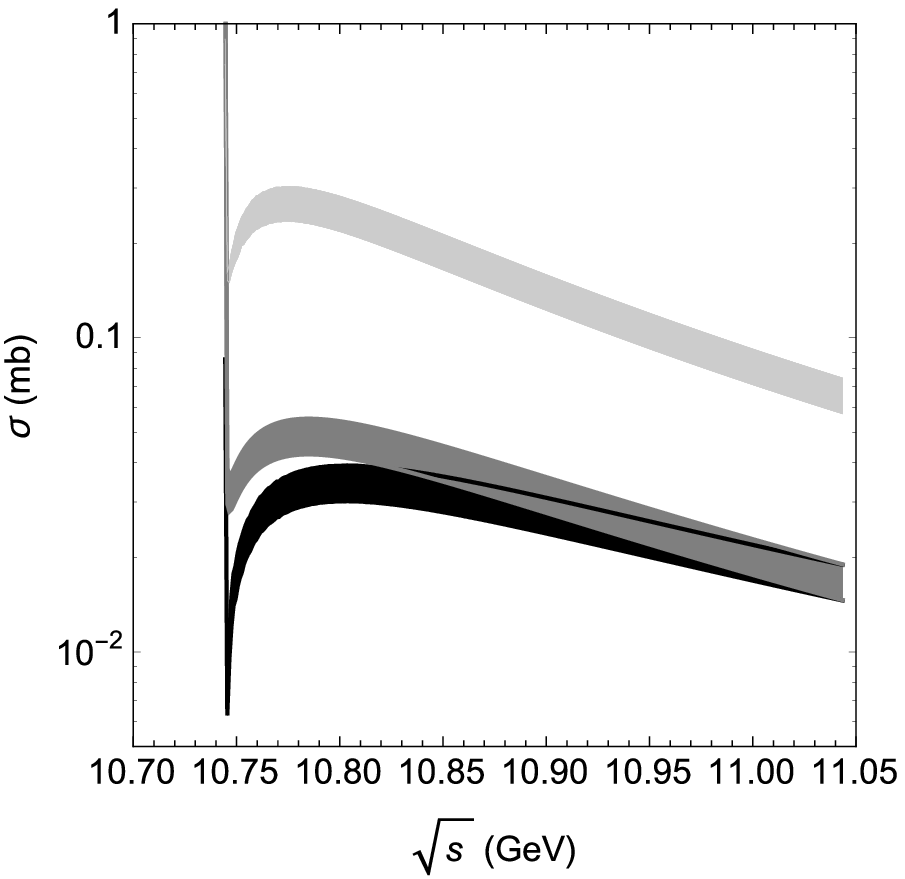} 
        \caption{Same as in Fig.~\ref{FIG6}, but with inclusion of form factors. }
    \label{FIG10}
\end{figure}



\begin{figure}[!ht]
    \centering
        \includegraphics[{width=8.0cm}]{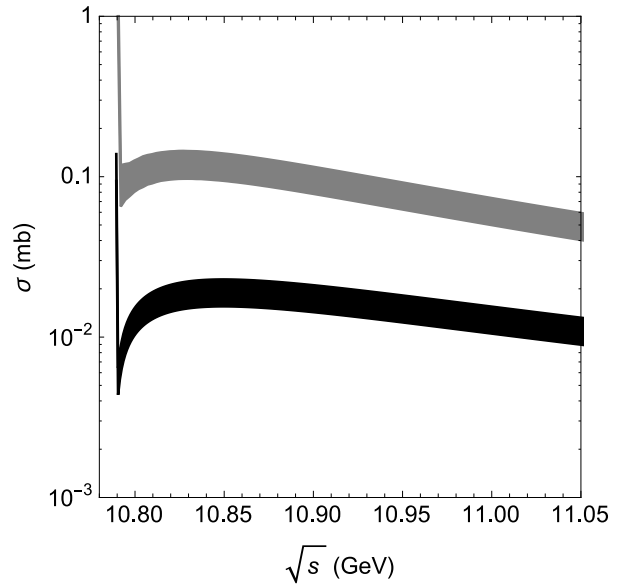} 
        \caption{Same as in Fig.~\ref{FIG7}, but with inclusion of form factors. }
    \label{FIG11}
\end{figure}


We can also include form factors in the vertices when evaluating the cross sections for the processes discussed above. Following  
~\cite{ChoLee,XProd},  we introduce a form factor of the type
\begin{eqnarray}
  F(\vec{q}) = \frac{\Lambda ^2}{\Lambda ^2 + \vec{q}^2}
  \label{ff}
\end{eqnarray} 
in the calculation of the cross sections for each of the vertices; $\Lambda $ is the cutoff and $\vec{q}$ the momentum
transfer in the CM frame [that is,  
$\vec{q}= (\vec{p}_{1CM} - \vec{p}_{3CM})$ for the $t$-channel, and $\vec{q}= (\vec{p}_{1CM} - \vec{p}_{4CM})$ for the $u$-channel].  

In Figs.~\ref{FIG8}-\ref{FIG11} we show the cross sections for the different reactions studied here when we include form factors in Eq.~(\ref{ff})
using $\Lambda = 2.0 $ GeV. As expected, the analysis done before remains qualitatively valid, but the magnitude of the cross sections suffers a reduction, 
especially at higher energies.



\section{Conclusions}

\label{Conclusions}



We have studied  the interactions between the $Z_b ^+(10610)$  (and  also $Z_b ^{\prime +}(10650)$) state and pions, in the processes 
$ \bar{B} B \rightarrow \pi Z_b $, $\bar{B}^* B \rightarrow \pi Z_b ^{(\prime)} $ and $\bar{B}^* B^* \rightarrow \pi Z_b ^{(\prime)} $ 
and their inverse reactions. We have obtained the amplitudes and cross sections related to these processes at leading order within the framework of $SU(4)$ effective Lagrangians. 

We have found that the $ Z_b ^{(\prime) +} $ production cross sections for the different final $\bar{B} B$, $\bar{B}^* B$, $\bar{B}^* B^*$  states are of the same order of magnitude. The same happens for the $ Z_b ^{(\prime) +} $ absorption cross sections.

But one of the main points here is that for reactions involving both $Z_b$ and $Z_b ^{\prime}$ states, the absorption cross sections are greater than the production cross sections, but still comparable with them. This fact may give a chance of significant survival probability of $Z_b ^{(\prime)}$ in heavy ion collisions. 

A similar result was found 
for the $X(3872)$ \cite{XProd}. However, whereas the $X$ absorption cross sections are, on average, about two orders of magnitude larger than the production ones, the $Z$'s absorption cross sections are only a factor about ten larger than the production ones. 
These significant differences of the cross sections imply that the $X$ and the $Z$'s are much more easily destroyed than produced in a hot hadronic medium, but 
the $Z$'s  have slightly better survival chances.


\acknowledgements

The authors would like to thank the Brazilian funding agencies CNPq, FAPESP and CAPES for financial support.


\end{document}